\documentclass[pra,onecolumn,floatfix,superscriptaddress]{revtex4}
\usepackage{graphicx}

\newcommand{\e}[1]{\ensuremath{\times 10^{#1}}} 
\newcommand{\up}{{\uparrow}}
\newcommand{\down}{{\downarrow}}

\newcommand{\ion}[2]{\mbox{$^{#2}$#1$^+$}}
\newcommand{\Ca}[1]{\ion{Ca}{#1}}

\newcommand{\lev}[2]{\mbox{#1$_{\mbox{\tiny$#2$}}$}}
\newcommand{\hfslev}[3]{\mbox{#1$^{\mbox{\tiny$#3$}}_{\mbox{\tiny$#2$}}$}}

\newcommand{\unit}[1]{\,\mbox{#1}}

\newcommand{\Hz}{\unit{Hz}}

\newcommand{\MHz}{\unit{MHz}}

\newcommand{\THz}{\unit{THz}}

\newcommand{\mW}{\unit{mW}}

\newcommand{\um}{\unit{$\mu$m}}

\newcommand{\s}{\unit{s}}

\newcommand{\us}{\unit{$\mu$s}}

\newcommand{\degree}{\mbox{$^{\circ}$}}

\newcommand{\ltish}{\protect\raisebox{-0.4ex}{$\,\stackrel{<}{\scriptstyle\sim}\,$}}

\newcommand{\bra}[1]{\mbox{$\left< #1 \right|$}}
\newcommand{\ket}[1]{\mbox{$\left| #1 \right>$}}

\begin{document}

\title{High-fidelity two-qubit quantum logic gates using trapped \Ca{43} ions}
\author{C. J. Ballance, T. P. Harty, N. M. Linke and D. M. Lucas \\
{\it Department of Physics, Clarendon Laboratory, Parks Road, Oxford OX1 3PU, U.K.}}
\date{20 June 2014}
\maketitle

The generation of entanglement is a fundamental resource for quantum technology, and trapped ions are one of the most promising systems for storage and manipulation of quantum information. Here we study the speed/fidelity trade-off for a two-qubit phase gate implemented in \Ca{43} hyperfine trapped-ion qubits. We characterize various error sources contributing to the measured fidelity, allowing us to account for errors due to single-qubit state preparation, rotation and measurement (each at the $\sim0.1\%$ level), and to identify the leading sources of error in the two-qubit entangling operation. We achieve gate fidelities ranging between $97.1(2)\%$ (for a gate time $t_g=3.8\mu$s) and $99.9(1)\%$ (for $t_g=100\mu$s), representing respectively the fastest and lowest-error two-qubit gates reported between trapped-ion qubits by nearly an order of magnitude in each case. 

We perform a two-qubit geometric phase gate in the $\sigma_z$ basis \cite{Leibfried2003}, where the qubits are stored in the \hfslev{S}{1/2}{4,+4} and \hfslev{S}{1/2}{3,+3} states of the ground hyperfine manifold of \Ca{43}. The two-qubit gate operation is implemented by a pair of Raman laser beams at a detuning $\Delta$ from the $\lev{4S}{1/2}\leftrightarrow\lev{4P}{1/2}$ transition. To vary $t_g$ we adjust $\Delta$ while holding the Raman beam intensity constant (at 5\mW\ per beam in a spot size of $w=27\um$); smaller $\Delta$ enables a faster gate, at the cost of increased error due to photon scattering~\cite{Ozeri2007}. The Raman difference frequency is $\delta = \nu_z + \delta_g$ where $\delta_g = 2/t_g$ and the axial trap frequency is $\nu_z = 1.95\MHz$. The Raman beams propagate at 45\degree\ to the trap $z$-axis, such that their wave-vector difference is along $z$. We cool both axial modes of the ions close to the ground state of motion by Raman sideband cooling; the centre-of-mass mode, rather than the stretch mode, is used to implement the gate to avoid coupling to the (uncooled) radial modes of the trap~\cite{Roos2008}.

We embed the phase gate within a single-qubit spin-echo sequence~\cite{Home2006}, which ideally produces the Bell state $\ket{\psi_+} = (\ket{\down\down}+\ket{\up\up})/\sqrt{2}$, and then use a further single-qubit rotation to measure the fidelity $F=\bra{\psi_+}\rho\ket{\psi_+}$ of the state $\rho$ obtained~\cite{Leibfried2003}. Thus the measured Bell state infidelity includes both errors due to the gate operation itself and errors in the single-qubit operations. We calibrate all single-qubit errors by independent experiments in order to extract the two-qubit gate error. The errors in the single-qubit operations are comparable to or smaller than the gate error over the parameter regime studied.

\begin{figure}[h]
\centering
\includegraphics[width=150mm]{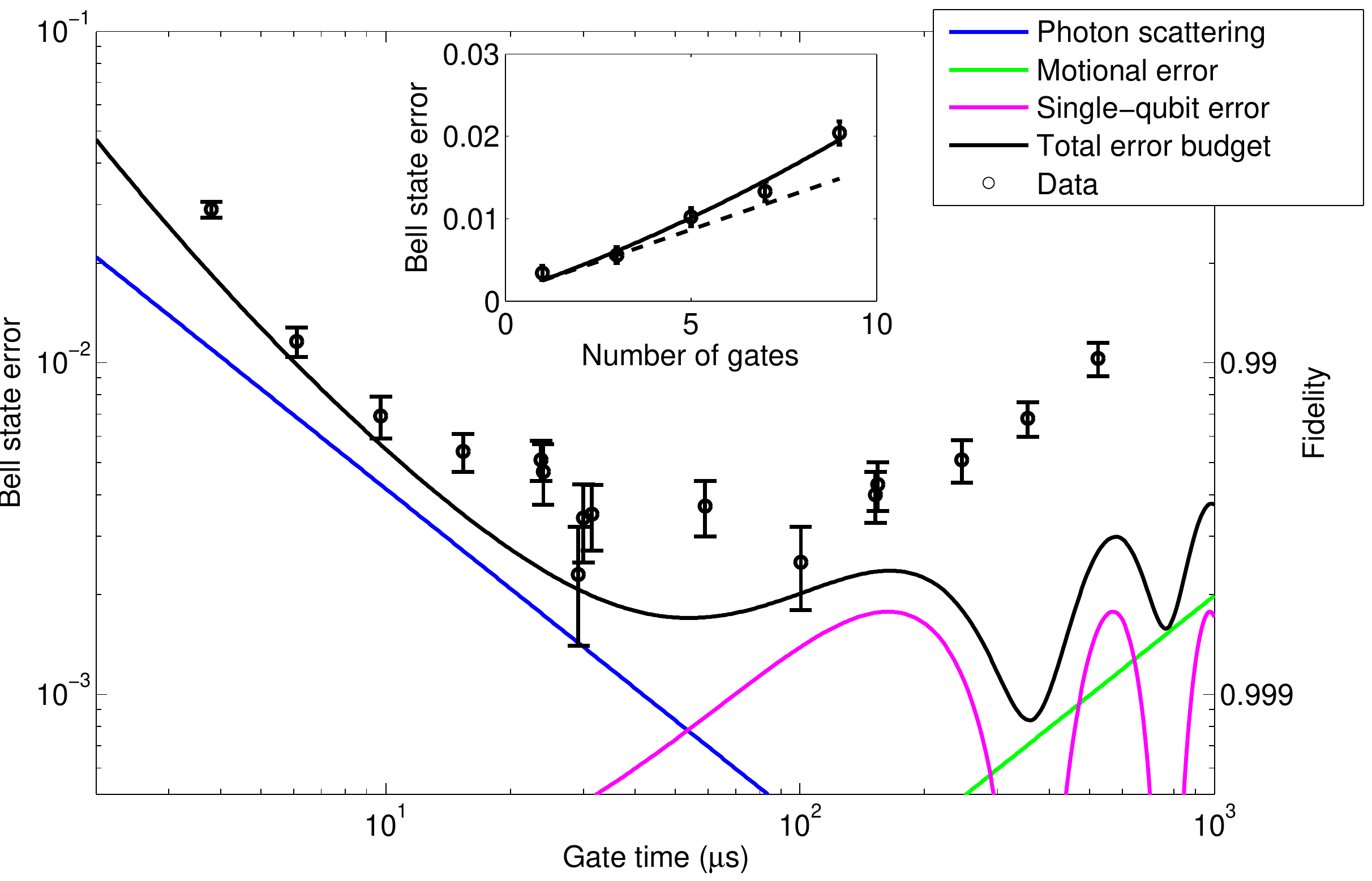}
\caption{Measured Bell state infidelity (data points) and error model (lines), plotted against the two-qubit gate duration $t_g$. Error bars are statistical. The three dominant contributions to the modelled total error (black line) are plotted. %
Inset: Bell state error vs (an odd) number of two-qubit phase gates, using $t_g=30\us$. The dashed line is the prediction from the error model of $15\e{-4}$ per gate, while the solid line is a quadratic fit which allows for a systematic error in the Raman beam intensity of 0.5\% (consistent with the precision to which it can be set).}
\label{F:allgates}
\end{figure}

Results are shown in figure~\ref{F:allgates}, where we have normalized for qubit readout errors (17\e{-4}). The data are in reasonable agreement with our error model for $t_g\ltish 200\us$; we attribute the excess error for longer $t_g$ to the effect of single-qubit dephasing errors (arising from the influence of magnetic field noise on the field-sensitive qubit states). The lowest gate error is found at $t_g=100\us$ (using $\Delta=-3.0\THz$), where the measured Bell state fidelity is $F=0.9975(7)$. For this run, the single-qubit error contribution is modelled to be 14\e{-4}, and we infer a gate error of $\epsilon_g=11(7)\e{-4}$. This is consistent with the known contributions to $\epsilon_g$ given in the table below. 

\begin{table}[h!]
\center
\begin{tabular}{|l|c|} \hline
Source & Error (\e{-4}) \\ \hline
Raman photon scattering   & $4$\\
motional heating and dephasing & $2$\\
motional temperature & $0.4$\\
Raman beam intensity systematic error  & $<1$ \\
off-resonant excitation & $<0.1$\\
\textbf{Total} & \textbf{7} \\ \hline
\end{tabular}
\label{T:budget}
\end{table}

We also performed multiple $t_g=30\us$ gates within a single spin-echo sequence (inset of figure~\ref{F:allgates}). As the single-qubit error is independent of the number of gates, this allows an independent estimate of the two-qubit gate error. 
The accumulated error after 9 gates implies an upper bound for the error per gate of 21\e{-4}, independent of all single-qubit and readout errors. This is consistent with the higher photon scattering error expected for this faster gate. Applying longer sequences of gates, for example for the purpose of randomized benchmarking~\cite{Gaebler2012}, would not be useful in this system as the measured error would be dominated by the effects of 50\Hz\ magnetic field noise rather than the accumulated gate errors. 

In separate experiments we have measured a coherence time of $T_2^*=50\s$ and a single-qubit gate error of $\approx 1\e{-6}$ using the ``atomic clock'' states of \Ca{43}, and shown the ability to map between the qubit we use here and the clock qubit with $<2\e{-4}$ error \cite{Harty2014}. Together with the two-qubit operations described here, this demonstrates -- in the same physical system -- a set of fundamental quantum logic operations with all errors approximately an order of magnitude below recent numerical estimates of the threshold required for fault-tolerant quantum computing \cite{Fowler2012}.

This work was funded by the U.K.\ Engineering and Physical Sciences Research Council (EPSRC), and will be described in more detail in a future paper. We are grateful to D.~Leibfried, D.~N.~Stacey, A.~M.~Steane and D.~J.~Wineland for helpful comments. 


\vspace{10ex}

\end{document}